\pdfoutput=1																										

\documentclass[preprint,3p,times,twocolumn]{elsarticle}				

\newcommand{\myeq}[3]{\vspace{#2} \begin{equation} \hspace{#1} #3 \end{equation} \vspace{0cm}}

\newcommand{\Op}{\mathcal{O}}

\newcommand{\no}{\noindent}
\newcommand{\vsp}[1]{\vspace{#1}}

\renewcommand{\bf}[1]{\textbf{\boldmath #1}}

\usepackage{amssymb}

\usepackage[figuresright]{rotating}

\begin{document}

\begin{frontmatter}




\title{Analysis of discrepancies in Dalitz plot parameters in $\eta\to 3\pi$ decay\tnoteref{label1}}
\tnotetext[label1]{This work was done in collaboration with J.~Novotn\'{y} and S.~Descotes-Genon. 
It was supported by the Center for Particle Physics (project no. LC 527) of the Ministry of Education of Czech Republic.}


\author{Mari\'{a}n Koles\'{a}r}

\address{Institute of Particle and Nuclear Physics, Faculty of Mathematics and Physics, Charles University, Prague}

\begin{abstract}
	We analyze the Dalitz plot parameters of $\eta$$\,\to$$3\pi$ decay in the framework of resummed chiral perturbation theory. This 	 
	approach allows us to keep the uncertainties in the NNLO and higher orders under better control and estimate their influence.   
	We cannot confirm the suspected discrepancy in the case of the charged decay parameter $b$, where even small uncertainties in higher 
	orders could accommodate the difference. On the other hand, we find the experimental value of the neutral decay parameter $\alpha$ 
	incompatible with an assumption of good convergence properties in the center of the Dalitz plot. We calculate $\pi\pi$ rescattering 
	bubble corrections up to three loops and show that these might explain the discrepancy, especially for a low value of the 
	pseudoscalar decay constant in the chiral limit. However, that could indicate a failure of convergence of the chiral series in this 
	channel already at low energies around 500MeV.
\end{abstract}


\end{frontmatter}


Chiral perturbation theory \cite{GL2,GL3} has a long history \cite{GLEta} of trying to explain
experimental data on the $\eta$$\,\to$$3\pi$ decay. As is clear from the decay rate
calculations, the theory converges really slowly for this decay channel.
While the latest NNLO calculations \cite{BijnEta} provide reasonable predictions 
for this quantity, experimental data are being gathered with increasing
precision in order to make more detailed analysis of the 
Dalitz plot distribution possible. The conventional Dalitz plot
parameters are defined as (we generally follow notations from \cite{BijnEta}):

\myeq{-0.5cm}{0cm}{
	\eta\to\pi^0\pi^+\pi^-:\,\ |A|^2 = A_0^2 (1+ay+by^2+dx^2+\dots)\quad }
\myeq{-0.5cm}{-0.5cm}{
	\eta\to 3\pi^0:\qquad |\overline{A}|^2 = \overline{A}_0^2(1+\alpha z+\dots)}	

\no where $x$$\,\sim$$\,u$$-$$t$, $y$$\,\sim$$\,s_0$$-$$s$, $z$$\,\sim$$\,x^2$+$y^2$ and 
$s_0$ is the Dalitz plot center $s_0=1/3(M_\eta^2$+$3M_\pi^2)$ .

Comparison of the recent experimental information with the NNLO $\chi$PT results
can be found in tables \ref{tab1}, \ref{tab2}.

\begin{table} \small
\begin{tabular}{|cc|c|}
	\hline \rule[-0.2cm]{0cm}{0.5cm} & & $\alpha$ \\
	\hline \rule[-0.2cm]{0cm}{0.5cm}  1998 & Crystal Barrel \cite{Cr.Barrel} & $-0.052\pm0.020$ \\
	\rule[-0.2cm]{0cm}{0.5cm} 2001 & Crystal Ball \cite{Cr.Ball} & $-0.031\pm0.004$ \\
	\rule[-0.2cm]{0cm}{0.5cm} 2007 & WASA at CELSIUS \cite{CELSIUS} & $-0.026\pm0.014$ \\
	\rule[-0.2cm]{0cm}{0.5cm} 2009 & WASA at COSY \cite{COSY} & $-0.027\pm0.009$ \\
	\rule[-0.2cm]{0cm}{0.5cm} 2009 & Cr.Ball at MAMI-B \cite{MAMI-B} & $-0.032\pm0.003$ \\
	\rule[-0.2cm]{0cm}{0.5cm} 2009 & Cr.Ball at MAMI-C \cite{MAMI-C} & $-0.0322\pm0.0025$ \\
	\rule[-0.2cm]{0cm}{0.5cm} 2010 & KLOE \cite{KLOE} & $-0.0301\pm0.0050$ \\
	\hline \rule[-0.2cm]{0cm}{0.5cm} 2007 & NNLO $\chi$PT \cite{BijnEta} & $+0.013\pm0.032$ \\
	\hline
\end{tabular}
	\caption{Recent experimental and $\chi$PT results for the neutral channel.}
	\label{tab1}
\end{table}\normalsize

\begin{table} \small
\begin{tabular}{|c|cc|c|}
	\hline \rule[-0.2cm]{0cm}{0.5cm}
			  & 1998 & 2008 & 2008 \\
	\rule[-0.2cm]{0cm}{0.5cm}
			  & Cr.Barrel \cite{Cr.Barrel+} & KLOE \cite{KLOE+} & NNLO $\chi$PT \cite{BijnEta} \\
	\hline \rule[-0.2cm]{0cm}{0.5cm}
			a & $-1.22\pm0.07$ & $-1.090\pm0.020$ & $-1.271\pm0.075$ \\
	\rule[-0.2cm]{0cm}{0.5cm}
			b & $0.22\pm0.11$ & $0.124\pm0.012$ & $0.394\pm0.102$ \\
	\rule[-0.2cm]{0cm}{0.5cm}
			d & $0.06\pm0.04$ & $0.057\pm0.017$ & $0.055\pm0.057$ \\
	\hline
\end{tabular}
	\caption{Recent experimental and $\chi$PT results for the charged channel.}
	\label{tab2}
\end{table}

\no As can be seen, the most
obvious discrepancy appears in the charged decay parameter $b$ and the neutral decay
parameter $\alpha$.

Other approaches were developed in order to model the amplitude better, namely dispersive
approaches \cite{Kambor,Anisovich,Lanz,Novotny} and non-relativistic effective field theory 
\cite{Bisseger,Gullstrom,Kubis}. These more or less
abandon strict equivalence to $\chi$PT and their success in reproducing a negative
sign for $\alpha$ can serve as a motivation to ask what is the culprit of the
failure of chiral perturbation theory to do so.

Our aim is thus not to produce another alternative method, but to try to understand whether
the theory, by which we mean $\chi$PT as a low energy representation of QCD, really does
have difficulties explaining the data and if so, try to identify the source of
the problem.

The starting point is the realization that the standard approach to $\chi$PT, as a usual 
treatment of perturbation series, implicitly assumes good convergence properties and hides
the uncertainties associated with a possible violation of this assumption. Error bars
are often not reported and it should be stressed that even the ones cited in the NNLO $\chi$PT
result above \cite{BijnEta} are not systematic uncertainties inherent in the theory, but rather 
fitting errors tied to numerical procedure peculiar to the method used by the authors.

There is a long standing suspicion that chiral perturbation theory might posses a slow
or irregular convergence in the case of the three quark flavor series \cite{Fuchs,Stern99}, the
$\eta$$\,\to$$3\pi$ decay rate might serve as a prime example. An alternative approach, dubbed resummed
$\chi$PT \cite{SternRes,Descotes}, was developed in order to express these assumptions in terms of parameters
and uncertainty bands. The procedure can be very shortly summarized in the following way:

\vsp{-0.1cm}
\begin{itemize}	
	\item[-]	standard $\chi$PT Lagrangian and power counting \vsp{-0.2cm}
	\item[-]	only expansions derived directly from the\\ generating functional trusted \vsp{-0.2cm}
	\item[-] 	explicitly to NLO, higher orders collected\\ in remainders\vsp{-0.2cm}
	\item[-]	remainders retained, treated as sources of error \vsp{-0.2cm}
	\item[-]	manipulations in non-perturbative algebraic way \vsp{-0.1cm}
\end{itemize}

Our calculation closely follows the procedure outlined in \cite{Kolesar}. What we present here is
only a brief excerpt, skipping all the details and concentrating only on the cases of the 
Dalitz plot parameters $b$ and $\alpha$. A more comprehensive work is in preparation \cite{prep}. 

Within the formalism, we start by expressing the charged decay amplitude in terms of the 4-point
Green functions $G_{ijkl}$. We compute at first order in isospin braking, 
in this case the amplitude takes the form

\myeq{-0.5cm}{0cm}{
	F_\pi^3F_{\eta}A(s,t,u)
		= G_{+-83}-\varepsilon_{\pi}G_{+-33}+\varepsilon_{\eta}G_{+-88} + \Delta^{(6)}_{G_D},\ }
		
\no where $\Delta^{(6)}_{G_D}$ is the direct higher order remainder to the complete 4-point Green function. 
The physical mixing angles to all chiral orders and first in isospin braking
can be expressed in terms of quadratic mixing terms of the generating functional to NLO
and related indirect remainders

\myeq{-0.5cm}{0cm}{
	\varepsilon_{\pi,\eta} = -\frac{F_{0}^{2}}{F_{\pi^0,\eta}^{2}}
		\frac{(\mathcal{M}_{38}^{(4)}+\Delta_{M_{38}}^{(6)}) - 									
		M_{\eta,\pi^0}^{2}(Z_{38}^{(4)}+\Delta _{Z_{38}}^{(6)})}
		{M_\eta^2-M_{\pi^0}^2}.}

\no In this approximation the neutral decay channel amplitude can be related to the charged
one as

\myeq{0cm}{0cm}{\overline{A}(s,t,u)=A(s,t,u)+A(t,u,s)+A(u,s,t).}

In accord with the method, strictly $\Op(p^2)$ parameters appear inside loops, while
physical quantities in outer legs. Due to the leading order masses in loops 
such a strictly derived amplitude has an
incorrect analytical structure, cuts and poles being in unphysical places. To account
for this, the amplitude is carefully modified using a NLO dispersive representation,
the procedure is described in detail in \cite{Kolesar}.   

The next step is the treatment of the low energy constants (LEC's). The leading order
ones are expressed in terms of convenient parameters

\myeq{-0.6cm}{0cm}{
	Z = \frac{F_0^2}{F_{\pi}^2},\ \
	X = \frac{2F_0^2B_0\hat{m}}{F_{\pi}^2M_{\pi}^2},\ \
	r = \frac{m_s}{\hat{m}},\ \
	R = \frac{(m_s-\hat{m})}{(m_d-m_u)},\ \ }
	
\no where $\hat{m}$=$(m_u+m_d)/2$. The standard approach tacitly assumes values of
$X$ and $Z$ close to one and $r$$\,\sim$25, which means that the leading order terms
should dominate the expansion. However, even the most recent standard $\chi$PT fit \cite{FitNew}
indicates a much lower values of $X$ and $Z$, thereby allowing for a possibility of
a non-standard scenario of spontaneous chiral symmetry braking (SB$\chi$S).

At next-to-leading order, the LEC's $L_4$-$L_8$ are algebraically reparametrized in terms
of pseudoscalat masses, decay constants and the free parameters $X$, $Z$ and $r$ using chiral expansions of 
two point Green functions, similarly to \cite{SternRes}. Because expansions are formally not truncated, 
each generates an unknown higher order remainder.

We don't have a similar procedure ready for $L_1$-$L_3$ at this point, therefore we collect
a set of standard $\chi$PT fits \cite{FitNew,Fit10,FitOld} and by taking their mean and spread, while 
ignoring the much smaller reported error bars, we obtain an estimate of their influence. As
will be shown in \cite{prep}, the results depend on these constants only very weakly. The error bands given
here include the estimated uncertainties in $L_1$-$L_3$.

\begin{figure}[t]
	\includegraphics[width=0.48\textwidth]{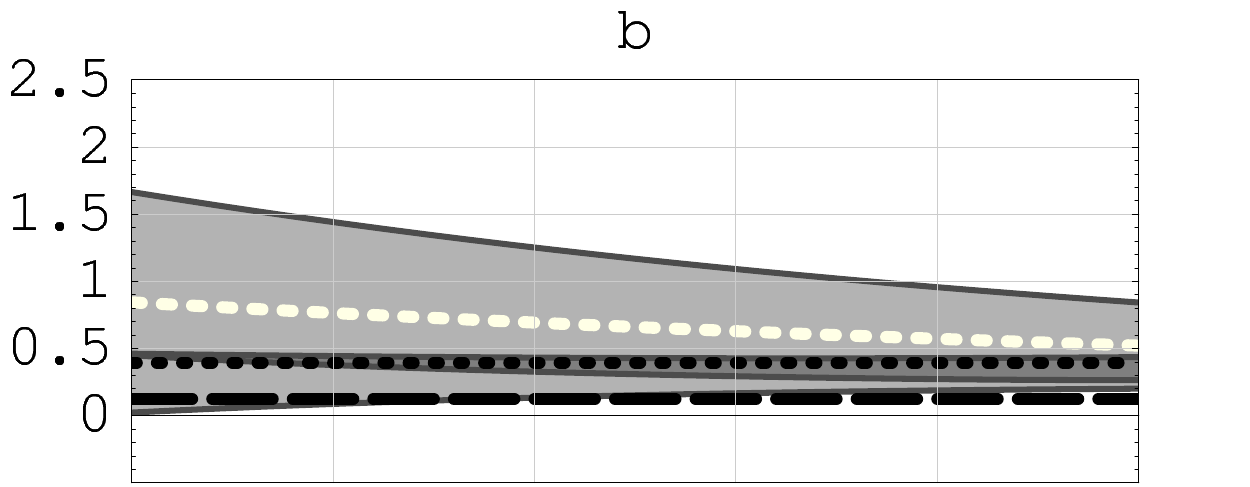}
	\includegraphics[width=0.475\textwidth]{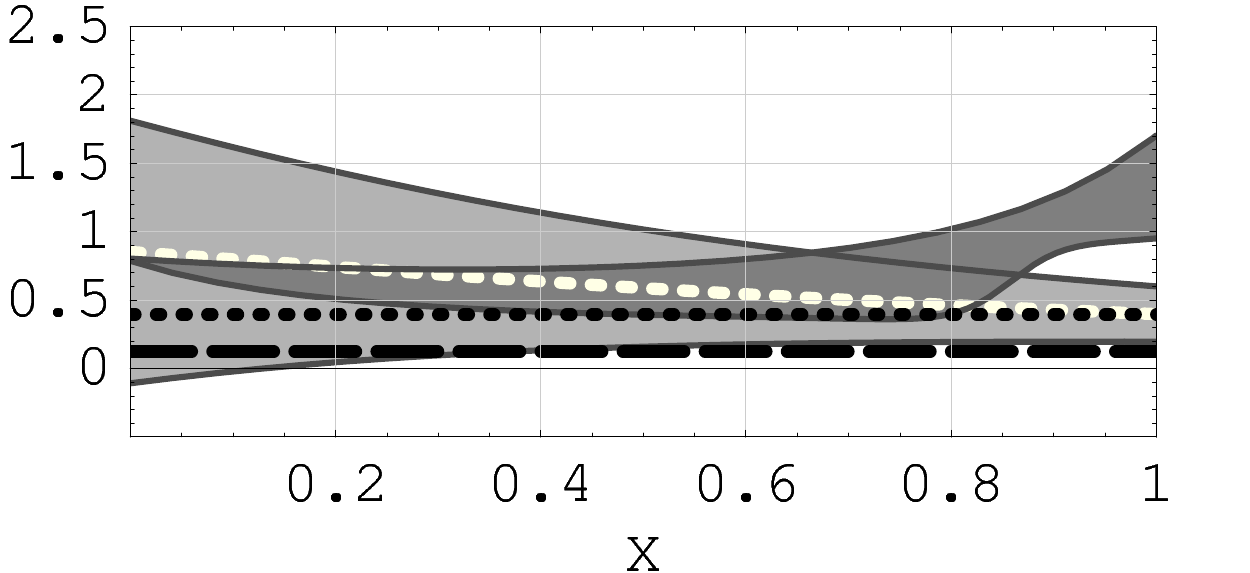}
	\caption{Charged decay parameter $b$: top $Z$=1, bottom $Z$=0.5\newline
	horizontal lines: dashed - KLOE measurment \cite{KLOE+}; 
	dotted - S$\chi$PT \cite{BijnEta} 
	light band - statistical remainder estimate; 
	dark band - result including $\pi\pi$ rescattering, 
	depending on scale $\mu$=0.5$\div$1GeV.}
	\label{fig1}
\end{figure}

\begin{figure}[t]
	\includegraphics[width=0.5\textwidth]{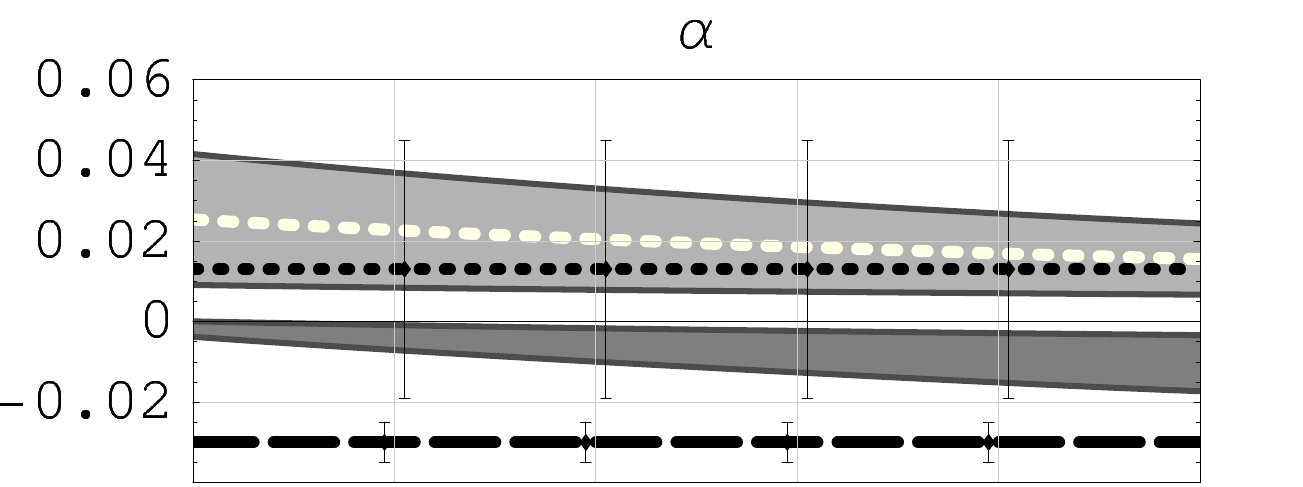}
	\includegraphics[width=0.5\textwidth]{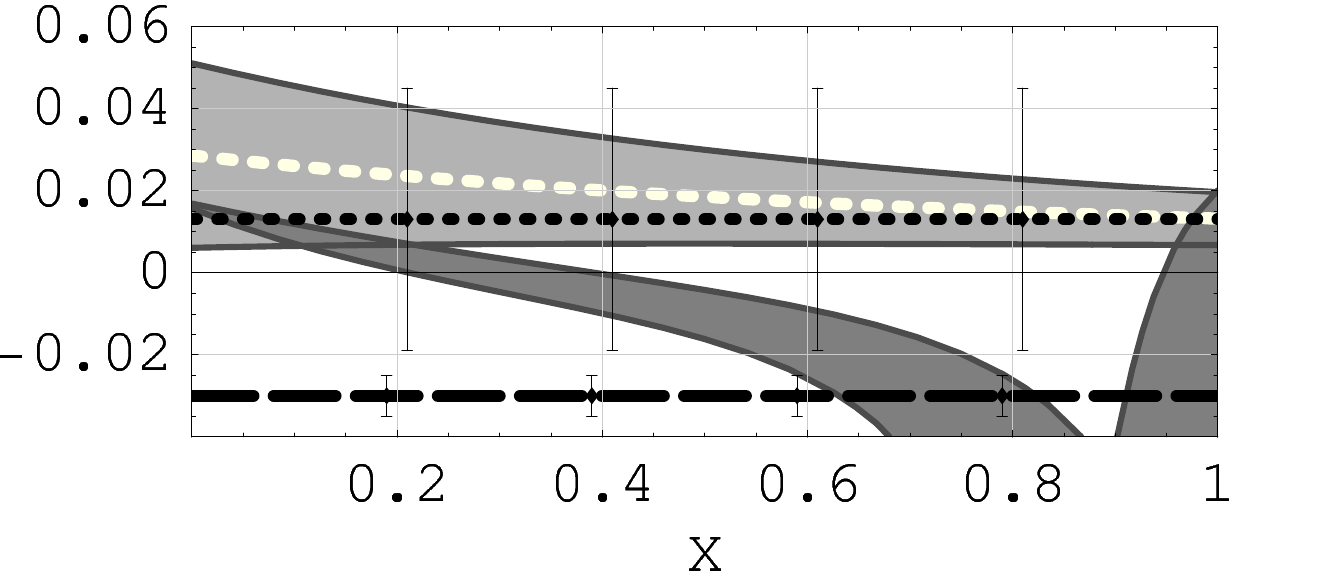}
	\caption{Neutral decay parameter $\alpha$: top $Z$=1, bottom $Z$=0.5\newline
	horizontal lines: dashed - KLOE measurment \cite{KLOE+}; 
	dotted - S$\chi$PT \cite{BijnEta} 
	light band - statistical remainder estimate; 
	dark band - result including $\pi\pi$ rescattering, 
	depending on scale $\mu$=0.5$\div$1GeV.}
	\label{fig2}
\end{figure}

The $O(p^6)$ and higher order LEC's, notorious for their abundance, are collected in
a relatively smaller number of higher order remainders. We also fix the $s$-quark mass
at $r$=25, motivated by lattice \cite{FLAG}, as its value is anyway connected with $X$ through
the Kaplan-Manohar ambiguity \cite{Kaplan}. We also investigated a low value $r$=15, but the
results does not radically alter the presented conclusions \cite{prep}. At last, because
at first order in isospin breaking the Dalitz plot parameters do not depend on
$R$,  we are left, besides the remainders, with two free parameters: $X$ and $Z$. These
effectively control the scenario of SB$\chi$S in our results.

Finally, the last step leading to numerical results is the estimate of the remainders.
We have a direct remainder to the 4-point Green function and eight indirect ones - three
related to both pseudoscalar masses and decay constants, two to mixing angles.
We use two approaches. The first one is based on general arguments about the convergence 
of the chiral series \cite{SternRes}, which leads to

\myeq{2cm}{-0.25cm}{\Delta_G^{(6)}\ \sim \pm 0.1 G,}

\no where $G$ here stands for any of our 2-point or 4-point Green functions,
which generate the remainders. This is in principle an assumption. Hence we test the
compatibility of this assumption of a reasonably good chiral convergence of trusted 
quantities with experimental data in a statistical sense. One peculiarity should be
noted though - Dalitz plot parameters are derivative quantities in terms
of the Mandelstam variables and even if the direct remainder had a small absolute value
around the center of the Dalitz plot, it could generate a large correction in derivatives. 
When making an expansion of the direct remainder around the Dalitz plot center, such
a large term would either grow to generate a large direct remainder in some other region, 
possibly unphysical, but still where $\chi$PT is assumed to converge well, or could be canceled 
by terms proportional to higher order derivatives, thus inducing some kind of fluctuation
in the amplitude. As we feel such a behavior warrants explanation, we include the 
absence of unusually large corrections in derivatives in our definition of good convergence 
properties and test for such a possibility as well.

The results for the statistical remainder estimate are depicted in fig.\ref{fig1} and \ref{fig2}.
Several conclusion can be made - in both cases the NNLO standard $\chi$PT result lies in our
uncertainty bands (the lighter ones), which is an important consistency check. 
For the charged decay parameter $b$, this is also true for the latest experimental measurement.
This means we cannot confirm any discrepancy, as even a small correction compatible with 
the assumption reasonable chiral convergence could explain the differences between theory 
and experiment. On the other hand, the situation is quite different in the case of the 
neutral decay parameter $\alpha$, where we can conclude, in accord with the NNLO S$\chi$PT 
result, that our definition of good chiral convergence is not compatible with the data. 
It is also notable that this is true for any value of the free parameters $X$ and $Z$ 
and hence an alternative scenario of SB$\chi$B, e.g. a small value of the quark condensate, 
is not the culprit here.
		
The framework of resummed $\chi$PT is well suited to include additional information about
higher orders from various sources \cite{Kolesar,FEta}. An independent estimate of the remainders
can on one side be an important check of the validity of the statistical remainder
estimate, as in the case of the parameter $b$, or could try to explain any deviances
from this assumption, which is our aim for $\alpha$.

While other estimates being in preparation \cite{prep}, here we employ a specific higher order
calculation, namely $n$-loop bubble contributions to final state $\pi\pi$ rescattering,
which generally take the form

\myeq{-0.6cm}{0cm}{G_{\pi\pi}^{(2n)} = 
		\sum_{k+l=1}^{n-1}\mathrm{P}_{kl}^{(n-k)}(s,t,u)\,\mu_\pi^k\,J^r_{\pi\pi}(s)^l
		+\ (t,u\ \mathrm{channels}),\ }

\no where $\mu_\pi$ and $J^r_{\pi\pi}(s)$ are the usual chiral logs and one loop
scalar functions \cite{GL3}, respectively. $\mathrm{P}_{kl}^{(m)}(s,t,u)$ is an
$m$-th order polynomial in the Mandelstam variables.
We compute up to $O(p^8)$, that means 3-loop diagrams with LO vertices and 2-loop ones
with one or two NLO counter terms. We stress that what we do is not a unitarization
procedure but a genuine $\chi$PT calculation. The motivation is that this is one of
the suspects that could explain the discrepancy in $\alpha$ \cite{Lanz,Kubis,Novotny}.
Terms with highest power in $s$ are of the form $\sim$$N \frac{s^n}{F_0^{2n}} J(s,m_\pi)^{n-1}$, 
where $N$ is a numerical factor. In our case we obtain $N^{(1-\mathrm{loop})}$=1/2, 
$N^{(2-\mathrm{loop})}$=4/3, $N^{(3-\mathrm{loop})}$=5/8, which implies that $N$ 
is not a suppression factor with LO vertices. Thus at 
$\left|\frac{s}{F_0^2} J(s,m_\pi)\right|$$\approx$1 convergence blows up, which a 
simple analysis can show is around $\sqrt{s}\approx$600-700MeV at $Z=$1 or
$\sqrt{s}\approx$400-500MeV at $Z$=0.5, where the renormalization scale 
runs through $\mu$=0.5$\div$1GeV.

The result can be seen as the dark bands in figures \ref{fig1} and \ref{fig2}. 
If they did not wander outside the light ones, it would indicate a
confirmation that the $\pi\pi$ rescattering bubble contributions agree with the
statistical remainder estimate. This is indeed true for $b$, except a small area 
of the parameter space. Once again the case of $\alpha$ is different and we can
see that $\pi\pi$ rescattering could generate a negative sign, especially in
a case of small value of the pseudoscalar decay constant in the chiral limit.

Of course, this is very far from a complete calculation to $O(p^8)$, which also 
expresses itself in dependence on the renormalization scale $\mu$. But the fact that the
scale dependence is quite benign could be also interpreted in the way that it is not 
unreasonable to consider the bubble $\pi\pi$ rescattering separately.

A further discussion can be made \cite{prep} and we will only summarize the results briefly -
the source of the large negative $\pi\pi$-rescattering contributions to $\alpha$ 
turns out to be where suspected, the terms leading in $s$. These generate a concavity in the
amplitude, which is measured precisely by $\alpha$ through the second derivative.
It is interesting to note that the $O(p^8)$ contributions to $\alpha$ are actually larger than
the $O(p^6)$ ones, and a quick check of the anticipated form of the $s$-leading terms
in even higher order contributions show that the next few orders can be expected to be large
and possibly negative as well. These terms do not induce a large correction 
to the direct remainder in the center of the Dalitz plot, but the concavity is 
connected with a quick failure of the convergence of chiral series in the suspected 
energy region. That is unphysical for the case of the $\eta$$\,\to$3$\pi$ decay, 
but could be an indication, if some other contribution do not cancel them,
of a breakdown of the chiral expansion at quite low energies. This could be
due to some higher energy structure present, for example a resonance.\\




\nocite{*}
\bibliographystyle{elsarticle-num}



\end{document}